\documentclass[pra,showpacs,twocolumn,floatfix]{revtex4}%
\usepackage{amsmath}
\usepackage{amssymb}
\usepackage{graphicx}
\usepackage{dcolumn}
\usepackage{bm}
\usepackage{latexsym}
\usepackage{graphicx}

\usepackage{natbib}

\begin{document}
 \newcommand{\ThGppp}{Th$^{3+}$}
 \newcommand{\ThRmppp}{$^{229m}$Th$^{3+}$}
 \newcommand{\ThRppp}{$^{229}$Th$^{3+}$}
 \newcommand{\ThSppp}{$^{232}$Th$^{3+}$}
 \newcommand{\bra}[1]{\langle #1|}
\newcommand{\ket}[1]{|#1\rangle}
\title{Observation of the 717 nm electric quadrupole transition in triply charged thorium}
\date{\today }
\author{A. G. Radnaev, C. J. Campbell, and A. Kuzmich}
\affiliation{School of Physics, Georgia Institute of Technology, Atlanta, Georgia
30332-0430}
\begin{abstract}
We excite the 717 nm electric quadrupole 6$D_{3/2} $ $\leftrightarrow$ 7$S_{1/2}$ transition in a laser-cooled \ThSppp ion crystal. The transition frequency and the lifetime of the metastable 7$S_{1/2}$ level are measured to be 417 845 964(30) MHz and 0.60(7) s, respectively. We subsequently employ the 7$S_{1/2}$ level to drive the ions with nanosecond-long 269 nm laser pulses into the 7$P_{1/2}$ level. The latter is connected to the 7$S_{1/2}$ electronic level within the $^{229}$Th nuclear isomer manifold by the strongest available electron-bridge transition, forming a basis for its laser excitation.
\end{abstract}
\maketitle

State-of-the-art trapped ion frequency standards utilize narrow optical transitions between valence electron orbitals within a single ion confined in an rf trap. These clocks are currently limited to fractional inaccuracies of $\sim 10^{-17}$ \cite{Rosenband2008,Tamm2009,Chwalla2009,Chou2010}.  A nuclear transition between two levels of identical electronic quantum numbers in a single $^{229}$Th ion could also be used as the basis for a clock \cite{Peik2003}. With suitably chosen states in the compound system (nuclear + electronic), all leading-order external-field clock shift mechanisms can be eliminated, leaving only higher-order, significantly smaller shifts.  This would relax technical requirements as compared to conventional optical clocks and potentially allow for fractional inaccuracies of $\sim 10^{-19}$ \cite{campbell2012}.

An important potential application of the $^{229}$Th nuclear clock is in the search for temporal variation of fundamental constants, particularly the fine structure constant $\alpha$. The most accurate laboratory searches for $\alpha$-variation to date are performed by measuring the ratio of atomic clock frequencies derived from different atomic systems over a long period of time. Because the two different systems have different sensitivities to $\alpha$-variation, the two clock frequencies would differentially shift, changing the frequency ratio. The keys to a sensitive probe are precise frequency measurement of the clock transitions and largely different sensitivities of the transitions to change in $\alpha$. In the case of the $^{229}$Th nuclear transition, a large enhancement over atomic systems in $\alpha$-variation sensitivity is predicted to exist due to near-cancellation of electromagnetic repulsion of the protons and of strong interactions among the nucleons \cite{Flambaum2006,Litvinova2009}.  This enhancement, combined with the clock transition's extreme accuracy potential, would lead to an improvement upon the current best measurement of $\alpha$-variation by possibly as many as five orders of magnitude when using state-of-the-art clock technology.

\begin{figure}[hb]
\includegraphics[width=3.1in]{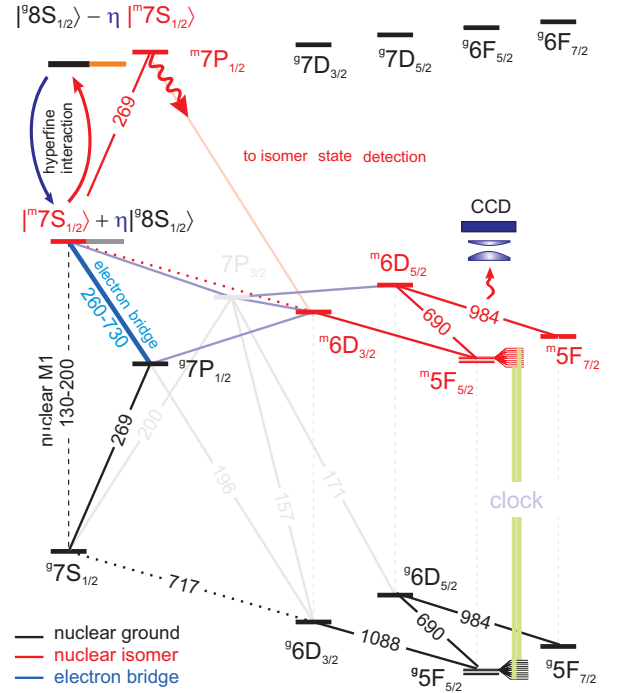}
\caption{Direct nuclear M1 and $\ket{^g7P_{1/2}} \rightarrow \ket{^m7S_{1/2}}$ electron bridge nuclear excitation schemes. The hyperfine interaction between the nucleus and the valence electron admixes electronic levels of the nuclear ground and isomer manifolds primarily through the $\ket{^g8S_{1/2}}$ and $\ket{^m7S_{1/2}}$ levels. This creates a $|^g7P_{1/2}\rangle \leftrightarrow |^m7S_{1/2}\rangle$ electron bridge with effective dipole moment $d_{EB}$ $\approx 2\times10^{-5} \, e \, a_0$ \cite{Porsev10}, which is substantially stronger than the bare nuclear M1 transition, and corresponds to a more convenient wavelength range.}
 \label{fig:EBbridge}
\end{figure}

\begin{figure*}[ht]
\includegraphics[scale=0.95]{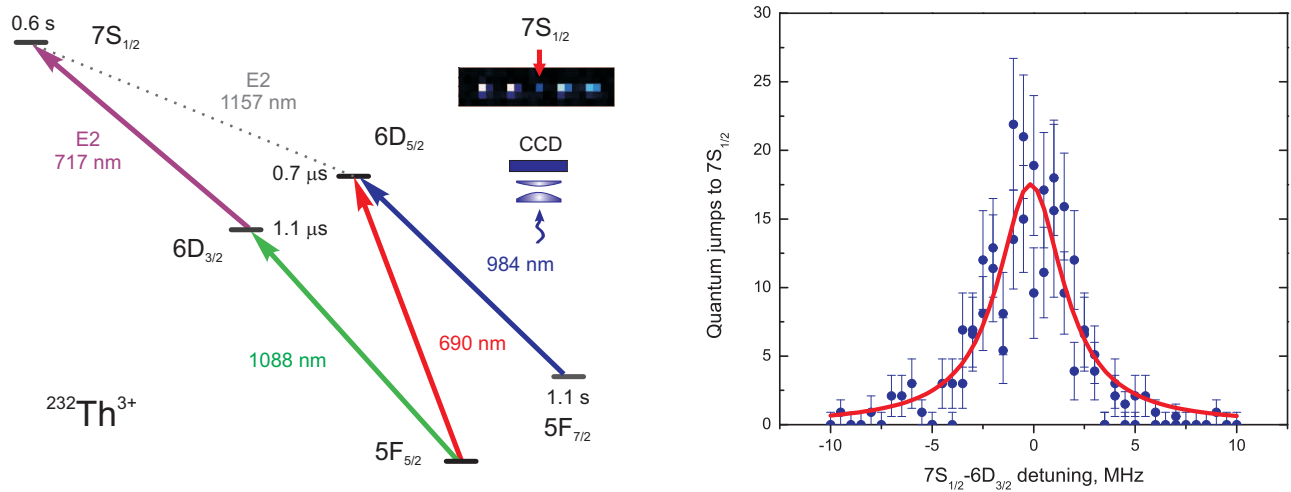}
\caption{(Left panel) Schematic of atomic levels used in the excitation of the 717 nm electric-quadrupole $6D_{3/2}\rightarrow7S_{1/2}$ transition. State detection is performed with 984 nm fluorescence measurements. The image shows one dark spot, surrounded by four bright ions. A dark spot typically corresponds to excitation of a $^{232}$Th$^{3+}$ ion to the metastable $7S_{1/2}$ level where 984 nm fluorescence is not produced. (Right panel) Measured profile of the 717 nm transition - the number of quantum jumps in 300 trials as a function of the 717 nm laser detuning.}
\label{fig:717_excitation}
\end{figure*}

 The most recent and precise published measurement of the $^{229}$Th isomer energy is 7.8(5) eV \cite{Beck07}.  In order to span $\pm\,3\,\sigma$ in the search for this nuclear level, direct optical excitation of the isomer in trapped cold ions may not be a viable method, given available UV sources and the large energy uncertainty.  Instead, the electron bridge (EB) process may be utilized, Fig. \ref{fig:EBbridge} \cite{Campbell09}.  In this case, hyperfine-induced mixing of the ground and isomer nuclear manifolds opens up electric-dipole transitions between the two.  Mixing is expected to be the strongest for the $S$-electronic states, as the electron probability density at the nucleus is highest.  For example, considering only the $7S_{1/2}$ and $8S_{1/2}$ electronic orbitals in first-order perturbation theory,
\begin{equation}
\nonumber
 \overline{|^m7S_{1/2}\rangle}  \approx  |^m7S_{1/2} \rangle
 + \frac{\langle ^g8S_{1/2}|H_{int}|^m7S_{1/2}\rangle}{ E_{^m7S_{1/2}} - E_{^g8S_{1/2}}} |^g8S_{1/2}\rangle,
\end{equation}
where $H_{int}$ is the electron-nucleus interaction Hamiltonian and $g(m)$ indicates the nuclear ground (isomer) level.
The $|^g8S_{1/2}\rangle$ admixture couples to the $|^g7P_{1/2}\rangle$ level via electric-dipole radiation of frequency $(E_{^m7S_{1/2}} - E_{^g7P_{1/2}})/\hbar$ (see Fig. \ref{fig:EBbridge}).  This shifts the spectral interrogation region from the challenging 130-200 nm range to the more promising 260 - 730 nm range.

A search for the nuclear isomer via the electron bridge in laser-cooled \ThRppp crystals will involve excitation to the $|^g7P_{1/2}\rangle$ level, followed by illumination with intense optical radiation which is tunable in the 260-730 nm range. Nuclear transition events will be manifested by interruptions in laser fluorescence on one of the laser cooling transitions. A chain of laser-cooled \ThRppp ions will provide near-unity detection efficiency of isomer state population, as well as strong localization for tight focusing of the isomer search light \cite{Campbell11}. In this paper, we demonstrate the first important step in this electron bridge-assisted isomer search by exciting $^{232}$Th$^{3+}$ ions to the $7P_{1/2}$ level via the three-step process: (i) $5F_{5/2}\rightarrow6D_{3/2}$ excitation at 1088 nm, (ii) $6D_{3/2}\rightarrow7S_{1/2}$ electric-quadrupole excitation at 717 nm, and (iii) $7S_{1/2}\rightarrow7P_{1/2}$ excitation at 269 nm.

In order to populate the $7S_{1/2}$ level, the $6D_{3/2}\rightarrow7S_{1/2}$ electric quadrupole transition at 717 nm is utilized. Because of limited quantity and high cost of the $^{229}$Th isotope, 717 nm and 269 nm spectroscopy is done here with the naturally occurring and less complicated $^{232}$Th.
\begin{figure*}[ht]
\includegraphics[width=7.0in]{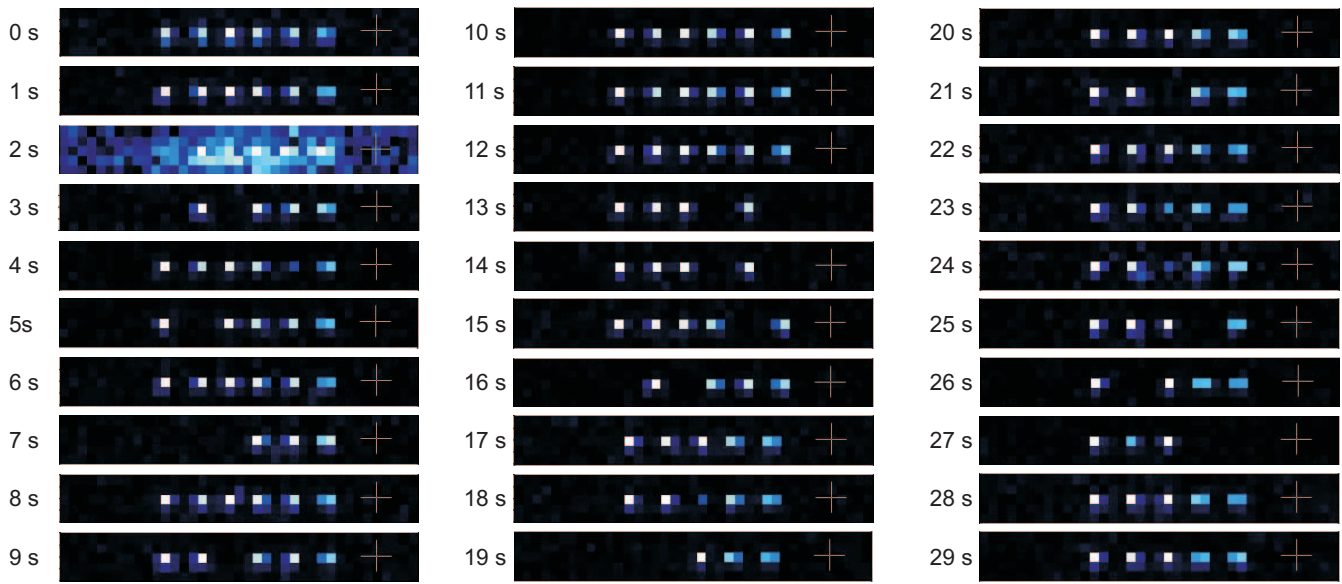}
\caption{A sequence of images of a \ThSppp chain, demonstrating observation of excitation to the metastable $7S_{1/2}$ level via the 717 nm electric quadrupole transition. Dark ions correspond to the excitation of the \ThSppp ion to the metastable $7S_{1/2}$ level, rather than a pollutant ion, inferred by the constant chain structure and expected sensitivity to the 717 nm laser frequency. A single-ion loss is evident between frames 16 s and 17 s, as the ion chain shortens by the corresponding length.}
\label{fig:717_video}
\end{figure*}
The quadrupole Rabi frequency between states $\ket{i}$ and $\ket{f}$ due to electric field of amplitude $\textbf{E}_0$ and wavevector $\textbf{k}$ is:
\begin{equation}
\nonumber
\Omega_Q = \frac{k_\alpha \bra{i}Q_{\alpha \beta}\ket{f}(E_0)_\beta}{\hbar}.
\end{equation}
The electric quadrupole matrix element $\bra{i}Q_{\alpha \beta}\ket{f}$ can be obtained using the Wigner-Eckart theorem and the reduced electric quadrupole moment $\bra{6D_{3/2}}|Q_{\alpha \beta}|\ket{7S_{1/2}}=7.0631 \, ea_0^2$ \cite{Safronova06}.
The 717 nm light is created by an external cavity diode laser (ECDL), where the frequency is stabilized to a transfer cavity as described in Ref. (\cite{Campbell11}).  An electro-optical phase modulator, driven by a voltage-controlled oscillator, is used to transfer $\sim$1 mW of optical power to ion resonance with the +1 order rf sideband.  The laser field is focused to a 110 $\mu$m diameter at the ions and corresponds to an electric quadrupole Rabi frequency $\Omega_Q \sim 100$ kHz. This value is sufficient to search the 2 GHz uncertainty region of the previously available indirect frequency measurement of 417 835(2) GHz that employed a gas discharge \cite{Klinkenberg88}.

The apparatus for creating, trapping, and laser cooling \ThGppp is described in Ref. \cite{Campbell11}. The trap rf frequency is increased to 8 MHz, which allows for operation without the aid of buffer gas. The electric quadrupole transition search protocol (Fig. \ref{fig:717_excitation}) consists of two main steps: (i) excitation to the metastable $7S_{1/2}$ level and (ii)  measurement of its population. For excitation of crystalized ions, the 690 nm repumping field is turned off and the atoms are optically pumped to the ground level with 984 nm light.  They are then excited to the $6D_{3/2}$ level with an axial 1088 nm field, corresponding to a $\sim$ 10 MHz Rabi frequency, and the frequency of the cw 717 nm excitation field is scanned. State detection is accomplished by measuring the population in the metastable $5F_{7/2}$ level, populated with the reintroduction of the 690 nm field. If an ion makes a transition to the $7S_{1/2}$ level, it is no longer resonant with the 984 nm detection field and does not scatter light, as illustrated in Fig. \ref{fig:717_video}. For the search, the interaction time is chosen to be 0.5 s, which is greater than the detection time of 0.1 s, but smaller than the theoretically predicted 0.59 s excited state lifetime \cite{Safronova06}.

The 717 nm laser frequency, spectrally broadened to $\approx 500$ MHz, is scanned in 500 MHz steps until the first transition is detected, after which the frequency is recorded. The broadening is achieved by modulating the control voltage of the voltage-controlled oscillator for the electro-optical phase modulator. The sequence is then modified to optimize the excitation rate with a shorter excitation time of 50 ms.  The rate of excitation to the $7S_{1/2}$ level is measured as a function of 717 nm field frequency, as shown in Fig. \ref{fig:717_excitation}. The $6D_{3/2}\rightarrow7S_{1/2}$ transition frequency in \ThSppp  is measured to be 417 845 964(30) MHz, as derived from a Lorentzian fit to the data and a corresponding wavemeter frequency measurement. The frequency uncertainty is dominated by the wavemeter inaccuracy. The measured transition frequency is $\approx$10 GHz higher than that given in Ref. \cite{Klinkenberg88}.

\begin{figure}[b]
\centering
\includegraphics[width=3.0in]{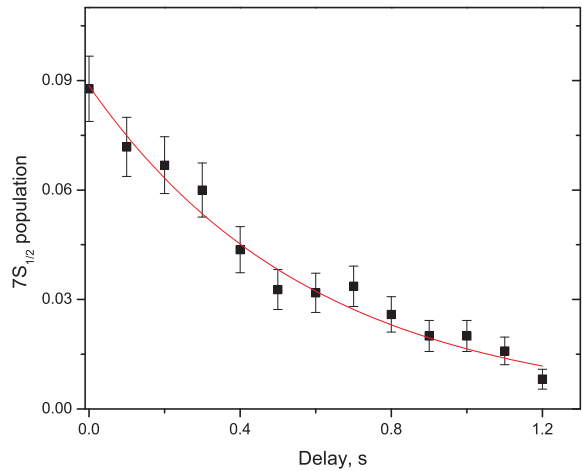}
\caption{Population of the 7$S_{1/2}$ level as a function of the measurement delay. The exponential fit gives a 1/e lifetime of 0.60(7) s.}
  \label{fig:f7s_lifeitme}
\end{figure}
\begin{figure*}[t]
\centering
\includegraphics[scale = 1.0]{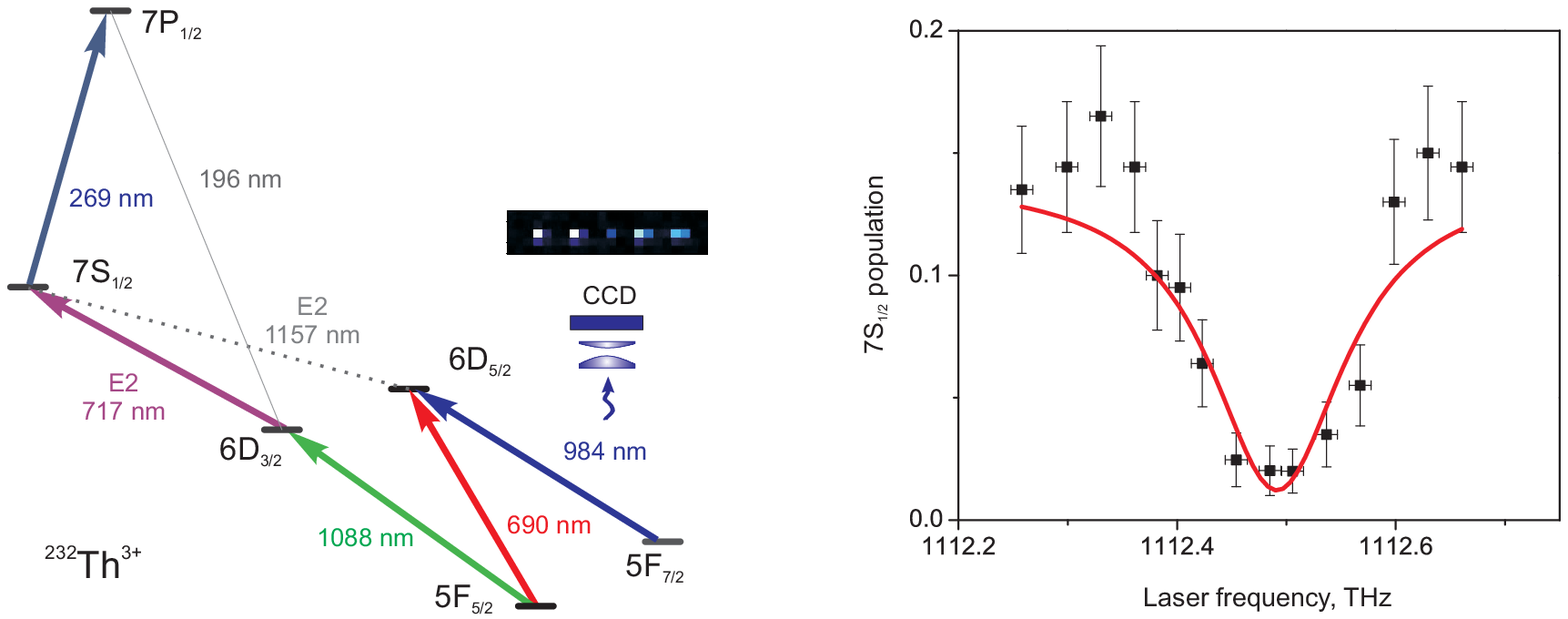}
\caption{(Left panel) Schematic of atomic levels used to excite the 269 nm transition. (Right panel) Measured population of the $7S_{1/2}$ level as a function of the 269 nm light frequency. Depopulation of the $7S_{1/2}$ level occurs when the 269 nm laser light is resonant with the $7S_{1/2}\leftrightarrow7P_{1/2}$ transition.}
\label{fig:269_excitation_232}
\end{figure*}
With the ability to efficiently transfer and observe population, this system is used to measure the lifetime of the 7$S_{1/2}$ level. The protocol starts with a 50 ms excitation pulse, followed by a variable delay and a 100 ms long fluorescence detection period. The total period of the sequence is set to 1 s to ensure return of the ions to the ground level. The population of the $7S_{1/2}$ level at various delays is shown in Fig. \ref{fig:f7s_lifeitme}. The exponential fit gives a lifetime of 0.60(7) s, in good agreement with the theoretically predicted value of 0.59 s \cite{Safronova06}. A significant systematic error can arise from residual $6D_{3/2}\leftrightarrow7S_{1/2}$ coupling caused by imperfect attenuation of the 717 nm field after excitation. We estimate the magnitude of this error to be much smaller than the statistical error of 0.07 s.

The $|^g 7P_{1/2}\rangle$ level is the initial state of the strongest electron-bridge transition,  $|^g 7P_{1/2}\rangle \leftrightarrow |^m 7S_{1/2}\rangle$. Because the $7S_{1/2}$ level can be populated with the electric quadrupole transition at 717 nm, the 269 nm $7S_{1/2}\rightarrow7P_{1/2}$ D$_1$ transition is a natural route toward population of the $7P_{1/2}$ level.

The laser light for excitation of this transition is obtained by a single-pass doubling of 539 nm light, generated by a home-built optical parametric oscillator (OPO). The OPO is based on a BBO crystal pumped at 10 Hz with 10 ns long, 355 nm pulses via the third harmonic of a flash-lamp-pumped YAG laser. The Littman configuration of the laser cavity allows for frequency tuning with a grating. The protocol for $7S_{1/2}\rightarrow7P_{1/2}$ spectroscopy is based on the sequence described above, with the addition of 269 nm light, as shown in Fig. \ref{fig:269_excitation_232}. The 717 nm laser excites ions into the metastable $7S_{1/2}$ level, revealed as dark spots in the ion chain of Fig. \ref{fig:717_video}. When the 269 nm field is resonant with the $7S_{1/2}\rightarrow7P_{1/2}$ transition, the atoms are strongly excited from the $7S_{1/2}$ metastable level for the duration of the pulse and the short ($\sim$1 ns) lifetime of the $7P_{1/2}$ level ensures rapid decay to the $|6D_{3/2}\rangle$ level. As the frequency of the 269 nm laser is scanned, the metastable state population is measured, as presented in Fig. \ref{fig:269_excitation_232}. The high contrast of the dip suggests nearly complete depopulation of the $7S_{1/2}$ level. The average intensity of the 269 nm laser is reduced to about 0.1 mW/cm${^2}$ of cw equivalent to minimize the linewidth of the spectroscopic signal, which is comparable to the expected linewidth of the 269 nm field.

With precise control of $7S_{1/2}$ and $7P_{1/2}$ excitation now realized in $^{232}$Th$^{3+}$, the next step will be to apply the techniques demonstrated here to the $^{229}$Th isotope. The transition frequencies between levels in $^{229}$Th$^{3+}$ are significantly shifted from those of the $^{232}$Th isotope due to large relative isotope shifts. The measured isotope shift for the $6D_{3/2}$ level is -9.856(10) GHz \cite{Campbell11}. For the $7S_{1/2}$ level, theory predicts an isotope shift coefficient of 146(4) GHz/fm$^2$ \cite{Berengut09}. Using the radius change $\langle r^2\rangle_{229} - \langle r^2\rangle_{232} = -0.300$ fm$^2$ based on measurements of Ref. \cite{Campbell11}, we obtain a -33.9(9) GHz expected relative isotope shift for the $6D_{3/2}\rightarrow7S_{1/2}$ transition.  The hyperfine splitting of the $7S_{1/2}$ level in \ThRppp is predicted to be 18(1) GHz. The hyperfine structure and relative isotope shift of the $7P_{1/2}$ level is expected to be covered by the $\sim$ 100 GHz linewidth of the 269 nm excitation field. This also importantly applies to the 269 nm transition within the nuclear isomer manifold.  The same field used for excitation to the electron bridge will also serve, once the isomer level is populated, to efficiently and rapidly transfer population from the $|^m 7S_{1/2} \rangle$ level to the $|^m 6D_{3/2} \rangle$ level for nuclear state detection.

It should be noted that the $|^g 7P_{1/2} \rangle$ electron bridge level can be populated with an alternative scheme - direct excitation from the ground state via the $5F_{5/2}\rightarrow7P_{1/2}$  two-photon transition at 329 nm. This scheme has the advantage of requiring fewer laser fields, however, the excitation pulses must be of much higher intensity. We have confirmed that \ThGppp  ions maintain a state of crystallization when exposed to high energy (100 mJ, 5 ns), tightly focused (40 $\mu$m diameter beam) pulses at 355 nm. Because the 329 nm excitation scheme requires light with much less intensity than this, the protocol should be feasible. Precise observation of the forbidden 717 nm transition, of which the transition frequency was not previously well known, also demonstrates that the current method of state detection is suitable for an efficient search of the electron bridge-assisted forbidden nuclear isomer transition in \ThRppp. This work was supported by the Office of Naval Research and the National Science Foundation.


\begin{thebibliography}{99}

\bibitem{Rosenband2008}
T. Rosenband {\it et al.}, Science {\bf 319}, 1808 (2008).

\bibitem{Tamm2009}
Chr. Tamm, S. Weyers, B. Lipphardt, and E. Peik, Phys. Rev. A {\bf 80}, 043403 (2009).

\bibitem{Chwalla2009} M. Chwalla {\it et al}., Phys. Rev. Lett. {\bf 102}, 023002 (2009).

\bibitem{Chou2010} C. W. Chou {\it et al.}, Phys. Rev. Lett. {\bf 104}, 070802 (2010).

\bibitem{Peik2003}
E. Peik and Chr. Tamm, Europhys. Lett. {\bf 61}, 181 (2003).

\bibitem{campbell2012}
C. J. Campbell {\it et al}., Phys. Rev. Lett. {\bf 108}, 120802 (2012).

\bibitem{Flambaum2006}
V. V. Flambaum, Phys. Rev. Lett. {\bf 97}, 092502 (2006).

\bibitem{Litvinova2009}
E. Litvinova, H. Feldmeier, J. Dobaczewski, and V. Flambaum, Phys. Rev. C {\bf 79}, 064303 (2009).

\bibitem{Beck07} B. R. Beck {\it et al.}, Phys. Rev. Lett. {\bf 98}, 142501 (2007); B. R. Beck {\it et al.}, LLNL-PROC-415170 (2009).


\bibitem{Campbell09}
C. J. Campbell {\it et al.}, Phys. Rev. Lett. {\bf 102}, 233004 (2009).

\bibitem{Porsev10} S.G. Porsev and V.V. Flambaum, Phys. Rev. A {\bf 81}, 032504 (2010).


\bibitem{Campbell11}
C. J. Campbell, A. G. Radnaev, and A. Kuzmich, Phys. Rev. Lett. {\bf 106}, 223001 (2011).


\bibitem{Klinkenberg88} P. Klinkenberg, Physica B+C \textbf{151}, 552 (1988).


\bibitem{Safronova06} U.I. Safronova, W. R. Johnson, and M. S. Safronova, Phys. Rev. A
  \textbf{74}, 042511 (2006).


\bibitem{Berengut09}
J. C. Berengut, V. A. Dzuba, V. V. Flambaum, and S. G. Porsev, Phys. Rev. Lett. {\bf 102}, 210801 (2009).


\end{thebibliography}
\end{document}